# A New Dynamic Model to Predict the Effects of Governmental Decisions on the Progress of the CoViD-19 Epidemic

Kamran Soltani[1], Ghader Rezazadeh[1, 2]

[1] Mechanical Engineering Department, Urmia, Iran

[2] South Ural State University, Chelyabinsk, Russian Federation

**Abstract**

*Background:* The outbreak of the new type of corona virus known as CoViD-19 in the entire world resulted in a large variation and impacts on worldwide interactions including, economics, social routine and etc. Unfortunately, it is still spreading and there is no certain vaccine yet, therefore, predicting CoViD-19 trend and understanding of its dynamic behavior may play a fundamental role in the prevention measures and control strategies. In the present paper, we have expressed a new look at the mathematical epidemic modeling based on the well-known compartmental models, to show the effects of the governmental decisions on the behavior of the epidemic in order to on-time effective actions.

*Methods:* We have established a novel mathematical model that considers various aspects of the spreading of the virus, including, the transmission based on being in the latent period, environment to human transmission, governmental decisions, and control measures. To accomplish this, a compartmental model with eight batches (sub-population groups) has been proposed and the simulation of the set of differential equations has been conducted to show the effects of the various involved parameters. Also, to achieve more accurate results and closer to reality, the coefficients of a system of differential equations containing transmission rates, death rates, recovery rates and etc. have been proposed by some new step-functions viewpoint.

*Results:* First of all, the efficiency of the proposed model has been shown for Iran and Italy, which completely denoted the flexibility of our model for predicting the epidemic progress and its moment behavior (using assumed logical model parameters). The model has shown that the reopening plans and governmental measures directly affect the number of active cases of the disease. Also, it has specified that even releasing a small portion of the population (about $2-3$ percent) can lead to a severe increase in active patients and consequently multiple waves in the disease progress. The effects of the healthcare capacities of the country have been obtained (quantitatively), which clearly specify the importance of this context. Control strategies including strict implementation of mitigation (reducing the transmission rates) and re-quarantine of some portion of population have been investigated and their efficiency has been shown.

*Future works:* This work is more and more efficient by undertaking statistical data for predicting and showing epidemic trends (Not just CoViD-19, but any other epidemic). So, our future work in this area may be dealing with the modeling of the other aspects of such

\* **Corresponding authors**: Kamran Soltani, Ghader Rezazadeh

Kamran Soltani (K. Soltani), kamransoltani16377@gmail.com, ORCID: 0000-0002-4479-2551,

Ghader Rezazadeh (G. Rezazadeh), g.rezazadeh@urmia.ac.ir, ORCID: 0000-0001-5243-3199.

epidemics and more certainty forecasting of the progression, of course by utilizing real statistical data.

# 1 Introduction

The outbreak of the new virus, known as SARS-CoV-2 or 2019-CoViD (severe acute respiratory syndrome corona virus 2), by the International Committee on Taxonomy of Viruses (ICTV), was started in December 2019 in Wuhan, China, when some incomprehensible symptoms were appeared in the patients [1,2]. The crisis of spreading this new type of virus (CoViD-19) had many notable effects on the typical trend worldwide such as economics, health, environmental and social, and unfortunately, it is still continued [3]. Therefore, fighting with this new virus to suppress and eliminate it is of particular importance.

In order to suppress the CoViD-19 disease, many preventive measures such as social distancing, quarantine, contact tracing, promotion wearing a face mask, the closing of high population density places such as schools, universities, offices, and etc. have been applied. However, the virus is progressing and still, there is no a certain and accessible vaccine to ward off the CoViD-19 epidemic, even if, it will be built soon, it will take a considerable amount of time to propagate in the entire of the world which is because of economic reasons and also maybe governments political reasons. During conflicts with the CoViD-19 epidemic, besides medical and biological efforts to find an effective vaccine, other scientific societies have a great responsibility to investigate other aspects of this pandemic including mathematical modeling of epidemics, biological systems modeling, and etc. [4–6].

Among, all of the above fields, the mathematical modeling can play a key role for predicting and determining the epidemic characteristics such as the pattern of the outbreak, the number of the people for example who infected (or the number of death cases), the threshold of the epidemic, the duration of the epidemic, effectiveness of the control measures and etc. [2,7]. So, mathematical modeling has a fundamental role in the perspective of governmental decisions and on-time actions to suppress and curb the spreading. The dynamical modeling of each epidemic has its own challenges that may lead to unexpected behavior and having some deviation from the real statistics information. The challenges of modeling of the epidemics such as CoViD-19 are because of their dependence on the many parameters and factors that they involved in the dynamical progress existing some vague medical issues, and regional reasons such as culture, being rich, and poor of the people and so on. In the early stage of most infectious diseases, the spreading dynamics is still unclear [8]. Mathematical modeling of the out breaking dynamics of the pandemic diseases including CoViD-19 implemented using a well-known family of the compartmental models [1,6,9], wherein the population is divided into subpopulation groups. The governing system of differential equations of the spread regarding the nature of the epidemic and its features are written and afterward, the flow of the transmission of the individuals from one compartment to another is obtained from solving a set of the governing differential equations [8]. There are various compartmental models with different structures; they are named based on the number and type of the compartments. At the beginning of the dynamical modeling of the epidemics, the number of boxes was limited to two or three compartments. The major classifications of the population in the compartmental models are susceptible-infectious (SI) [10], susceptible-infectious-susceptible (SIS) [10], susceptible-infectious-recovery or susceptible-infectious-removed (both known as SIR) [6,11–13], susceptible-infected-recovered-dead (SIRD) [14,15], susceptible-exposed-infectious-recovery (SEIR) [1,6,11,16] as well as more generalizations of the SEIR model [2,6,9,17,18], that can be found in the literature.

In the present study, a new mathematical modeling of epidemic disease for CoViD-19 is conducted. Governmental decisions (actions) and environmental disease transmission are modeled by a new approach. In order to closer predicting of the epidemic pattern some coefficients including infection rates, death rate, are proposed time-dependent, and the results for the model under time-varying coefficients is obtained. To show the efficiency and flexibility of the proposed model in the outbreak prediction, by considering logical model parameters the patterns of the spreading for Iran and Italy are obtained. During this, governmental decisions, change of transmission rates with time and etc. are employed. Multiple waves of the epidemic, healthcare capacity effects, environmental transmission effects and etc. are predicted and the suppressing controlling measures are shown. The results of the analysis are summarized as a proposal to make effective measures by the government and society.

## 2 Mathematical Modeling

As mentioned before, many compartmental models have been developed to model infectious diseases. Most of them are a generalization of the SEIR model and adding some new classes of the population try to consider as many involved parameters and conditions as possible. Here, too we follow the same process and consequently we consider a compartmental epidemic model by adding several new batches of the peoples. Meanwhile, unlike most of the existing models, the presented model takes into account infection of the susceptible individuals from the people who are in the latent period of the disease. This issue is very important especially for diseases like CoViD-19, which has a notable latent period. Based on the proposed compartmental model (figure (1)), the entire population is divided into eight batches. First of all, the people who are resistant to the disease, known as insusceptible ($IS$), these individuals are subtracted from the entire population. The other class is known as susceptible ($S$), which they able to get the disease by contact with both of exposed ($E$) and infected ($I$) individuals, the exposed people are those who they do not have any disease symptom yet, and they are in the incubation period (here it is considered equal to the latent period) of the disease, where the term of "incubation period" referred to a duration of time that the symptoms of the disease have not appeared yet (especially for cases like CoViD-19) [17,19]. Therefore, the susceptible individuals become infected at a rate $\beta$ by contact with the infected individuals and at a rate $\alpha$ by contact with the asymptomatic people (Exposed individuals). Where the coefficients $\alpha$ and $\beta$ are the transmission rates.

For cases similar to CoViD-19 the main cause of the contagion is associated with the connecting between susceptible and exposed people (hence $\alpha \gg \beta$). In addition, susceptible individuals may become infected by contact with a polluted environment (at the rate of $\xi$). It is assumed that the infection rate by the polluted environment to humans is a function of exposed people ($\xi = \xi(E, disinfection\ measures)$) as well as disinfection measures. The individuals from the exposed group are transferred to the infectious people at an appearance rate, $\varepsilon$, equal to $1/incubation\ period$. The Presented model uses an added isolated

(hospitalized), $H$, compartment and a quarantined ($S_q$) compartment to include the effects of the isolation of the infectious individuals and also the quarantine of the susceptible people. It is assumed that the isolated persons cannot infect new individuals from the susceptible people, also, we suppose that they are added to the isolation compartment by a fraction, $q$, of the infectious individuals. There are two possibilities for other remaining individuals in the infectious group, one, those who are recovered after passing from the corresponding infectious period of the disease and one, who has died. We suppose that, the recovery rate of the infectious individuals is $\gamma$ where it is equal to $1/infectios\ period$, as well as, the death rate (or mortality rate) from the infectious and quarantined people are $d_i$ and $d_q$, respectively. It should be mentioned that a fraction of the individuals from isolation compartment has been considered as $q_r$ which they are recovered. Accordingly, the set of differential equations of the dynamical modeling of the CoViD-19 can be written as follows:

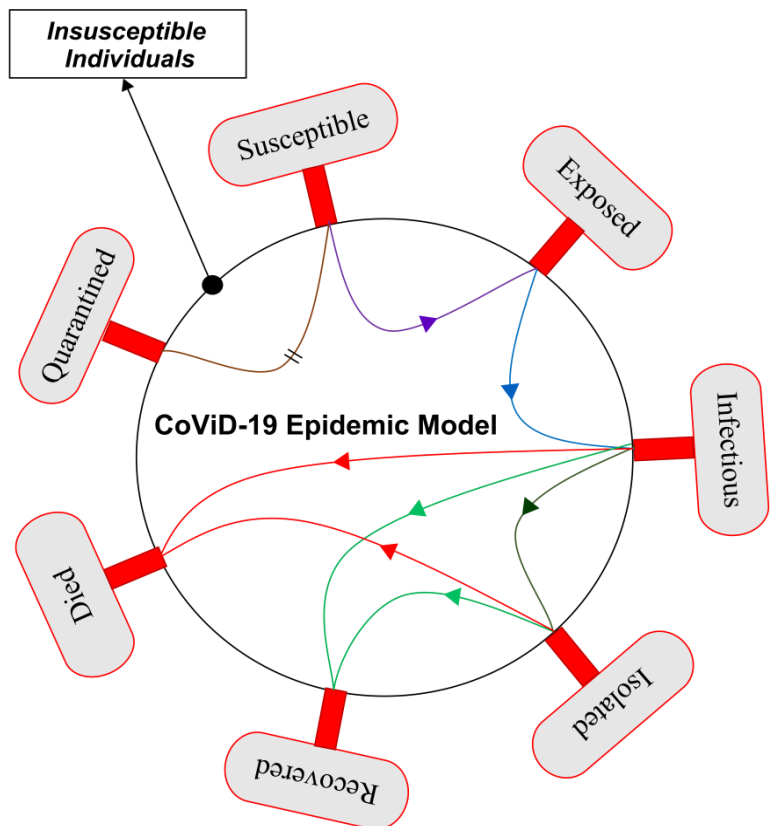

*Fig 1: The proposed SEIR epidemiological model of CoViD-19 disease.*

$$\Omega 1: \frac{dS(t)}{dt} = \frac{dN(t)}{dt} - (\beta(t)I + \alpha(t)E(t))S(t) - \xi(E)S(t) + Gd(t)$$

$$\Omega 2: \frac{dE(t)}{dt} = (\beta(t)I + \alpha(t)E(t))S(t) - \varepsilon E(t) + \xi(E)S(t)$$

$$\Omega 3: \frac{dI(t)}{dt} = \varepsilon E(t) - \gamma I(t) - d_i(t)I(t) - qI(t)$$

$$\Omega 4: \frac{dH(t)}{dt} = qI(t) - q_r H(t) - d_h(t)H(t)$$

$$\Omega 5: \frac{dR(t)}{dt} = \gamma I(t) + q_r H(t)$$

$$\Omega 6: \frac{dD(t)}{dt} = d_i(t)I(t) + d_h(t)H(t) \qquad (1)$$

$$\Omega 7: \frac{dS_q(t)}{dt} = -Gd(t)$$

$$\Omega 8: \frac{dN(t)}{dt} = \frac{IS(t)}{dt} + \frac{S(t)}{dt} + \frac{E(t)}{dt} + \frac{I(t)}{dt} + \frac{H(t)}{dt} + \frac{R(t)}{dt} + \frac{D(t)}{dt} + \frac{dS_q(t)}{dt}$$

$$\Omega 9: Gd(t) = \sum_{i=1}^{n} \bar{\delta}_i \exp(-\frac{(t-T_i)^2}{2\mu_i^2}) \quad , \quad \bar{\delta}_i = \frac{\Delta \bar{S}_0^i}{(2\pi\mu_i^2)^{0.5}}$$

In the proposed model, $Gd(t)$, is a function of applying the effects of the governmental decisions on the quarantine of the susceptible individuals at different times. More details about the $Gd(t)$ transmission rates will be discussed in the next sections. $T_i$, is the decision time. The number of the susceptible individuals also changes by the natural birth and mortality rates in the society, so this effect is modeled as $dN/dt$ in the susceptible compartment. A notable assumption of the presented model is that the all population is not subjected to the disease, directly, that means the quarantine people are immune from the infection to the disease until they will be subjected to disease, directly, (by contacting with other exposed/infected people or polluted public places). The summation of the eight groups is the total population of the society as follows:

$$IS(t) + S(t) + E(t) + I(t) + H(t) + R(t) + D(t) + S_q(t) = N(t) \qquad (2)$$

It should be noted that, in the system of differential equations (1), Ω8 is known as mass conservation equation and is obtained from the derivative of equation (2). Table (1) provides a brief definition of each compartment, in order to better understand their distinction.

**Table 2**: The definitions of the compartments and the individuals

| Compartment | Explanation |
| --- | --- |
| *Insusceptible (IS)* | ■ The people who are resistant against the disease (immune against the disease) |
| *Susceptible (S)* | ■ The people who are subjected to the disease directly (by their especial work, job or duty), these people can be infected. |
| *Quarantined ($S_q$)* | ■ The individuals that they are not subjected to the disease directly these individuals cannot be infected unless the quarantine is removed (e.g. the community of students because of closing schools and universities), |
| *Exposed (E)* | ■ Infected with the disease but without any typical symptoms (in incubation period) |
| *Infectious (I)* | ■ Infected by virus with general symptoms of the disease |
| *Isolated (H)* | ■ Individuals who have been isolated (hospitalized people) |
| *Recovered (R)* | ■ People who have been healed and are not infected again |
| *Death (D)* | ■ People who have died! |

It is worth mentioning that applying mathematical models on the systems in the real-world (physical, social, economic, biological and etc.) will be valid only under its utilized hypothesis and assumptions [6]. Hence, this research is the same as other mathematical models that will have a bit of deviation from the clinical information and formal statistics. Table (2) shows an overview of the main assumptions which are considered in the presented compartmental model. Also, there are many noise and perturbations in the real statistical data that is because of the many parameters that affect the behavior of the system in reality (the nature of our world). The practical mathematical description of the physical phenomenon can predict only a general behavior of the system by considering the most effective parameters.

**Table 2**: An overview of main assumptions, used in the present analysis.

| Related part | Assumptions |
| --- | --- |
| Variables | • The variables used in the presented model are assumed to be continuous in time |
| Birth and natural deaths | • It is assumed that the number of newborns and died individuals are neglected relative to the total population, so the population is remained constant by a good approximation. |
| Male and female | • In the present model there are no distinguished differences between male and female, although, the rate of the infection of the men and women are different [6]. |
| Age effects | • have been not considered. |
| Geopolitical factor | • have been not considered. |
| Recovery of cases | • It is assumed that recovered people will no longer be infected. |

**2-1 Governmental Decision effects**

As mentioned briefly before, large scale governmental decisions such as closing or opening of public communities, workplaces and etc. lead more individuals to be subjected with the virus. This is because of increasing the number of direct contacts or the rising infection rates, suddenly. In the present study, this phenomenon is modeled as an impact on the variations of the people subjected to the disease (susceptible individuals). The first reason is the novelty (the main one) of the presented paper, which considers the effects of the quarantine on the transmission of the susceptible people into exposed and then infectious individuals. That means that, during an epidemic outbreak, when the government decides for opening (closing) of a portion of the population, it leads to suddenly increasing (decreasing) of the susceptible compartment. Therefore, the governmental decisions are considered as a sum of Gaussian functions (Ω9) at the start of the performing of a decision. The justification of this selection is because of the step variations of the susceptible individuals, hence its variations (derivative respect to time) is impulsive. Also, when a change occurs, it is not expected that the model parameters including coefficients of transmission rates remain constant, therefore, in the perspective of the compatibility between the coefficients we assume that the other rates in the set of equations change. One of the main factors in determining the prevalence of the epidemic disease like CoViD-19 is the transmission rates ($\beta$ and $\alpha$) that they denote the rate of transferring individuals from the compartment susceptible to the compartment of exposure.

The infection rate depends on many various parameters such as culture, healthcare possibilities, the amount of economic well-being of the people, educational infrastructure, and government assistance for people during for example quarantine and etc. Therefore, it seems logical that we assume the infection rate coefficients are time-dependent ($\beta(t)$ and $\alpha(t)$). Any decision on a large scale in a country can impress the transmission rates and other existent coefficients. But as we will show later, even decreasing the transmission rates after the reopening plans, the breaking of lockdown may lead to an increase in the infectious people dramatically. Accordingly, Governmental measures control the progress of the outbreak directly, for example, in Iran at the beginning of the spreading of CoViD-19 some preventative measures such as banned festival celebration, closing Friday Prayers, disinfection of the public places and paths, closing holy shrines, closing schools, universities and offices were applied which reduced the rate of the spreading the CoViD-19 (Fig (2)) [20]. As can be understood from figure (2), after the reopening of some workplaces and jobs on around May 3, the number of infectious cases began to rise again.

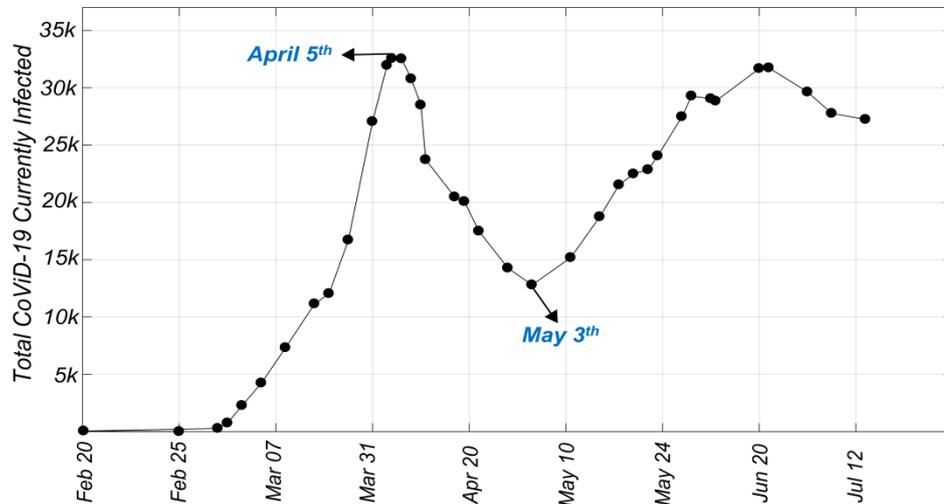

**Fig 2**: *Active infectious cases in Iran from February 20$^{th}$ to July 14$^{th}$* [20].

The idea of the modeling of the variations of the infection rates with time seems to be interesting for in the investigation of the behavior of the epidemics for example repeating of the waves of the disease. For this purpose various functions have been used in the literature to describe the infection rate as a function of time for other epidemics [21,22]. In the present study, we provide a sum of step functions for transmission rate coefficients during a certain period. Also, it is assumed that the infection rates change after the governmental decisions, since the government may apply the necessities of the personal preventative measures such as wearing a face mask, and so on. Following stepwise function is used for infection rates and the rate of fatality:

$$\beta(t) = \beta_0 u_0(t) + (\beta_1 - \beta_0) u_1(t) + \ldots + (\beta_N - \beta_{N-1}) u_N(t) \left(1 - \frac{P(t)}{N}\right)^\kappa \quad (3)$$

$$= \left(1 - \frac{P(t)}{N}\right)^\kappa \cdot \left(\beta_0 u_0(t) + \sum_{i=1}^{N} (\beta_i - \beta_{i-1}) u_i(t)\right)$$

$$\alpha(t) = \alpha_0 u_0(t) + (\alpha_1 - \alpha_0)u_1(t) + \ldots + (\alpha_N - \alpha_{N-1})u_N(t)\left[1 - \frac{P(t)}{N}\right]^\kappa \quad (4)$$

$$= \left[1 - \frac{P(t)}{N}\right]^\kappa \cdot \left(\alpha_0 u_0(t) + \sum_{i=1}^{N}(\alpha_i - \alpha_{i-1})u_i(t)\right)$$

$$d(t) = (d_0 u_0(t) + (d_1 - d_0)u_1(t) + \ldots + (d_M - d_{M-1})u_M(t)) \quad (5)$$

$$= \left(d_0 u_0(t) + \sum_{i=1}^{M}(d_i - d_{i-1})u_i(t)\right) \quad \text{Death Rate}$$

where $u_i(t)$, is a shifted unit step function that is introduced as follows:

$$u_i(t) = u(t - \tau_i) \quad (6)$$

the coefficients $\beta_i$, $\alpha_i$ and $d_i$ are the mean value of transmission rates and death rates over the time interval between $\tau_{i-1}$ and $\tau_i$, respectively. Also $P(t)$ is a term known as "Perception of risk" that directly depends on the mortality rates ($d_i$ and $d_h$) [21]. The parameter of the perception of risk, depending on the nature of the epidemic, is increased when more individuals are died (increasing death rate) and it is reduced naturally. For the presented model for CoViD-19 the variations rate of the perception of risk can be expressed by the following equation:

$$\frac{dP(t)}{dt} = \delta D(t) + e_m P(t) - \eta P(t) \quad (7)$$

the constant $\kappa$ is a parameter that determines the strength of the response. Also $\delta^{-1}$, $\eta^{-1}$ and $e_m^{-1}$ are the meantime of the transmission of the infectiousness to the deaths, the average duration of the impact of death cases on the perception of the public and the duration of the media effects on the general mentality, respectively, where both parameters to be estimated.

**2-2 Epidemic model Threshold**

One of the key parameters that determine the threshold of the outbreak is called basic reproduction number, $\mathcal{R}_0$ (or basic reproductive ratio) which is defined as the number of produced secondary infected cases in a completely susceptible population [8,23]. When $\mathcal{R}_0 < 1$, the spread is suppressed rapidly (decreases with time), by other words, the disease is not an epidemic. For case $\mathcal{R}_0 > 1$ the disease is spread, and it is continued until all the individuals in the society will be infected [24]. Noting to the definition, there are two general methods for determining $\mathcal{R}_0$, one using a statistical clinical-related data fitting and other using a mathematical model of the epidemic (known as Next generation method in literature) [24]. The mathematical calculation of the $\mathcal{R}_0$ is more popular when we have a compartmental model since it gives a closed form expression for $\mathcal{R}_0$ as a function of model parameters. For determining the epidemic progress in the absence of the $Gd(t)$ we calculate the basic reproductive ratio for the provided model. According to the proposed model, (when there is no $Gd(t)$ term in the equations), a unique disease-free equilibrium (DFE) can be obtained as follows:

$$X_0 = (S^0, S_q^0, E^0, I^0, H^0, R^0, D^0) = (f\bar{N}, (1-f)\bar{N}, 0, 0, 0, 0, 0) \quad (8)$$

where $f$, is the initial fraction of the susceptible population ($\bar{N}$) that are subjected to the disease, directly. At first, the next-generation matrix (NGM) should be formed, for this purpose, compartments $E$, $I$ and $H$ are considered as disease compartments and other as non-disease compartments. More details about formation of NGM can be found in Refs [19,25,26]. So, according to the set of equations (1), the matrix of new infections, $F$, and the matrix of the transition of individuals from disease compartments, $V$, are expressed as the following:

$$F = \begin{bmatrix} \left[\alpha f\bar{N} + f\bar{N}\cdot\frac{\partial \xi}{\partial E}\bigg|_{E=0}\right] & \beta f\bar{N} & 0 \\ 0 & 0 & 0 \\ 0 & 0 & 0 \end{bmatrix} \quad and \quad V = \begin{bmatrix} \varepsilon & 0 & 0 \\ -\varepsilon & \gamma + q + d_i & 0 \\ 0 & -q & q_r + d_h \end{bmatrix} \quad (9)$$

therefore, the NGM, $FV^{-1}$, can be obtained as following relationship:

$$FV^{-1} = \begin{bmatrix} \dfrac{\left[\alpha f\bar{N} + f\bar{N}\cdot\frac{\partial \xi}{\partial E}\big|_{E=0}\right]}{\varepsilon} + \dfrac{\beta f\bar{N}}{(\gamma + q + d_i)} & \beta f\bar{N}(\gamma + q + d_i) & 0 \\ 0 & 0 & 0 \\ 0 & 0 & 0 \end{bmatrix} \quad (10)$$

Finally, $\mathcal{R}_0$, is calculated by determining the spectral radius (matrix eigenvalue with the largest magnitude) of NGM, we have:

$$\mathcal{R}_0 = \rho(FV^{-1}) = \frac{\alpha f\bar{N}}{\varepsilon} + \frac{\beta f\bar{N}}{(\gamma + q + d_i)} + \frac{f\bar{N}}{\varepsilon}\cdot\frac{\partial \xi}{\partial E}\bigg|_{E=0} \quad (11)$$

at which the first two terms are related to the new infection by susceptible-exposed and susceptible-infected contacts, respectively. Also, the third term measures the infected ones by a polluted environment, as can be seen disinfection can directly affect the progress of the epidemic [19].

### 2-3 Environmental pollution effects

Regarding the lifespan of the CoVid-19 on the various surfaces, it can be concluded that, the environmental transmission is a notable source in the spreading of the virus. To take into account this subject our model added a new term of interacting susceptible people with the polluted environment that as said before is a function of the rate of disinfection by people and related organs and the number of the exposed individuals. In this paper, we assumed that the variations of the environmental transmission rate with the rate of disinfection are negligible relative to the variation with the number of exposed people. The main source of the pollution-related transmission is the public places; therefore, the $\xi$ function has been selected in a way that is proportional to the number of exposed individuals exponentially. Of course, by defining a saturation point a better approximation is obtained. Accordingly, the following function is considered for the relativity of the environmental transmission rate with the exposed people:

$$\xi(E) = \xi_0(1 - e^{-\xi_{00}E}) \qquad (12)$$

where $\xi_0$ and $\xi_{00}$ are the positive coefficients to adjustment of the function for real data. Also, the $\xi_0$ is the saturation level of environmental pollution. Finally, the basic reproductive number, $\mathcal{R}_0$, comes to the following form:

$$\mathcal{R}_0 = \rho(FV^{-1}) = \frac{(\alpha + \xi_0\xi_{00})f\bar{N}}{\varepsilon} + \frac{\beta f\bar{N}}{(\gamma + q + d_i)} \qquad (13)$$

In the present study, it is assumed that the CoViD-19 epidemic is near its saturation point in terms of environmental transmission, so, $e^{-\xi_{00}E} \approx 0$. However, statistical data can be useful for predicting as better of the environmental-related infection.

## 4 Result and Discussion

### 4-1 Simulation

As can be understood, the set of differential equations (1) is a nonlinear system (caused by terms $\beta SI$ and $\alpha SE$), so the exact solution maybe is not practically possible. For solving this system, familiar numerical methods can be used. In the present study, the $4^{th}$-order Runge-Kutta method is used for solving the governing nonlinear equations. According to the physics of the problem, the set of differential equations (1) needs a series of initial conditions for the involved variables. In the present analysis, it is assumed that, at the beginning of the outbreak, a fraction of susceptible people population ($f$), are subjected to the disease, also, only one person has been exposed to the virus (Table (3)). It should be noted that, some parameters have a relatively constant value or they change in a relatively specific interval which depend on the structure and the nature of the virus, such as incubation period and infectious period (for these coefficients the mean value is considered), in the other hands, there are some coefficients, that they change with many external parameters. These coefficients often vary from one community to another (e.g. infection rates and the environmental polluted coefficient). The values of parameters and coefficients have been brought in Table (3) for Iran and Italy. In the present analysis, due to the severe lack of statistical data, some parameters have been assumed, however, assumed parameters are relatively logical (according to the available data) but they need accurate estimation in the future works. Methods such as maximum-likelihood parameter estimation and least square are well known in the area of the data estimation [27,28]. Details of these coefficients have been shown in Table (4). More accurate simulations by including extensive statistical research can be conducted in future works (this paper alone cannot include this wide range of data estimation).

### 4-2 model Validation: Iran and Italy

In order to show the flexibility and efficiency of our proposed model to cover the existing data and verifying the model, the behaviors of the spreading CoViD-19 in Iran and Italy have been derived for a relatively long term of the epidemic (Figure (3)) using some logical model parameters. As denoted before, accurate data analysis and advanced statistical works by incorporating a verity of the related parts (Including ministry of health and hospital statistical centers etc.) are needed to more accurately predict the outbreak in order to apply on-time of prevention strategies.

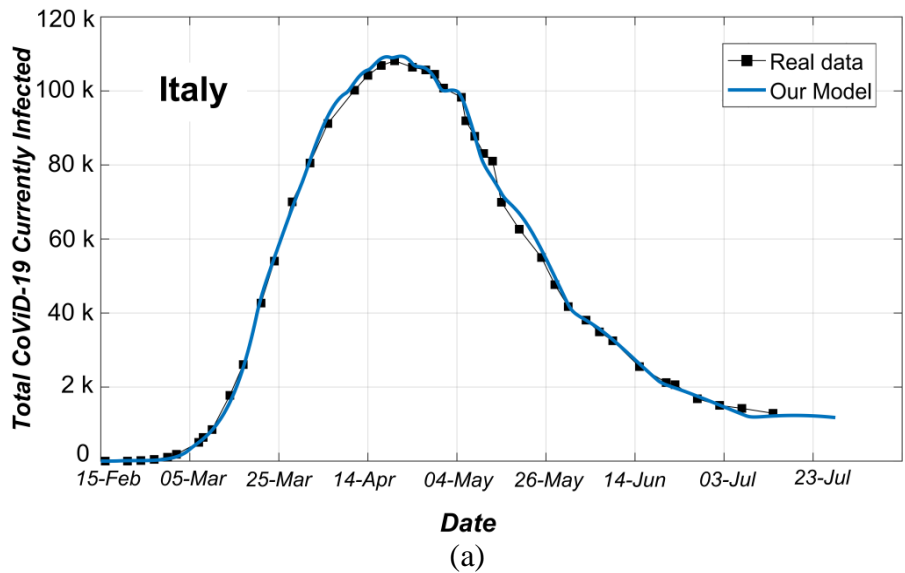

(a)

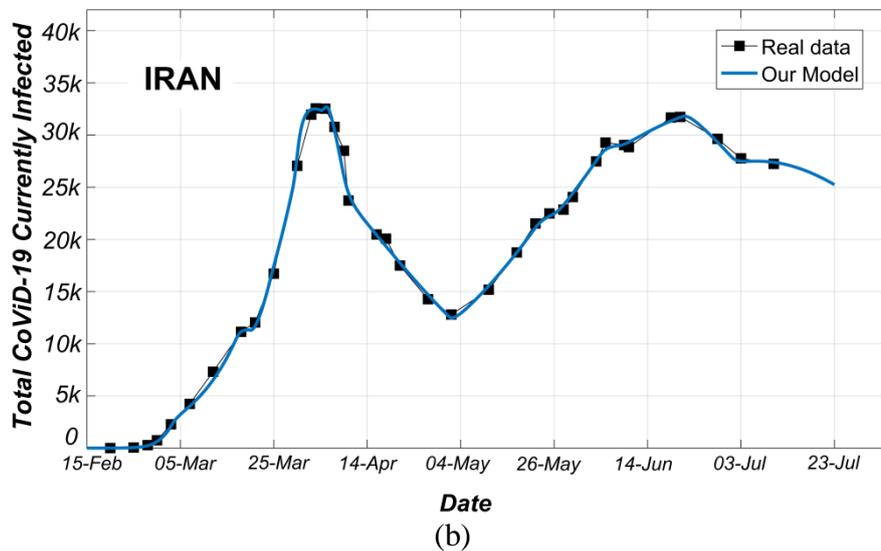

(b)

**Fig 3**: *Model validation for active infected cases in a) Italy and b) IRAN using proposed dynamical model* [20].

As Figure (3) illustrates, our model is able to well predict disease progression, both in terms of trend and quantity. Note that for deriving the results for both countries of Italy and Iran relatively reasonable coefficients and governmental measures have been considered. For example, for Italy, it has been assumed that at the beginning of the epidemic, since some people have been concerned (because of the media effects and the news of the world) they have begun to respect self and family isolation (at home generally). Also, they have removed unnecessary commuting, so it has been assumed that only 90 % (0.9$N$) of the population are subjected to the virus directly. Also, the exposure to susceptible transmission rate ($\alpha$) is time-dependent and varies tangible, but the infectious to susceptible infection rate ($\beta$) has a slower variation with time, since it often occurs in hospitals and treatment centers, therefore a mean constant value has been selected. On the other hand, a few days after the entrance of the virus to the country the government may have declared the quarantine, so a considerable portion of the population was prevented from the direct exposure to the disease (for Italy it is assumed 60 % of the population). In the progress of the epidemic CoViD-19 the government may decide to remove the lockdown of some special cortexes (due to their essential jobs) gradually (about 4 % for the case of Italy until Jul 14). A similar trend can be imagined for epidemic progress in Iran.

Table 3: Summarized values of the parameters and variable used in the analysis.

| Parameter & Discerption | Notation | Value | Remark |
|---|---|---|---|
| **Population and initial Conditions (Feb $20^{th}$)** | | | |
| Total population of Italy | $N^{(Italy)}$ | 60.45 million | |
| Total population of Iran | $N^{(Iran)}$ | 80.39 million | |
| Initial susceptible people | $S_0$ | $0.7 N^{(Italy)}$ | Assumed |
| | | $0.9 N^{(IRAN)}$ | Assumed |
| Initial exposed people | $E_0$ | 1 | |
| Initial infectious cases | $I_0$ | 0 | |
| Initial death cases | $D_0$ | 0 | |
| Initial quarantined people | $Q_0$ | 0 | |
| Initial recovered people | $R_0$ | 0 | |
| Initial quarantined people | $S_{q0}$ | From mass conservation | |
| Initial insusceptible people | $IS_0$ | 0.0 | |
| **Coefficients** | | | |
| Incubation (or Latent) period | $\varepsilon^{-1}$ | $\cong$ 5.1 day, [29] | |
| Infectious period | $\gamma^{-1}$ | $\cong$ 7 day, [29] | |
| Fraction of isolation of active cases | $q$ | 75 % Italy | Assumed |
| | | 70 % IRAN | |
| Mean death rate | $d$ | 0.00043 Italy | |
| | | 0.0006 IRAN | |
| Mean period of quarantined of active cases | $q_r^{-1}$ | 14 day [8] | |
| Mean environmental infection rate | $\xi$ | $4.3 \times 10^{-8}$ Italy | Assumed |
| | | $5.0 \times 10^{-8}$ IRAN | |
| Death perception of risk coefficient | $\delta$ | 1/8 day $^{(-1)}$ | Assumed |
| Medial effect coefficient | $e_m$ | 1/3 day $^{(-1)}$ | Assumed |
| Coefficient of naturally decay of risk | $\eta$ | 2/3 day $^{(-1)}$ | Assumed |
| Risk strength response | $\kappa$ | 500 | |
| Infectious to exposed transmission rate | $\beta$ | $5 \times 10^{-11}$ Italy | Assumed |
| | | $0.5 \times 10^{-11}$ IRAN | Assumed |

Table 4: The considered parameters for cases of Italy and IRAN.

| | IRAN | | | | | | | | | | | | | | |
|---|---|---|---|---|---|---|---|---|---|---|---|---|---|---|---|
| Infection rate $\times 10^{-8}$ ($day^{-1}$) | $\alpha_0$ | $\alpha_1$ | $\alpha_2$ | $\alpha_3$ | $\alpha_4$ | $\alpha_5$ | $\alpha_6$ | $\alpha_7$ | $\alpha_8$ | $\alpha_9$ | $\alpha_{10}$ | $\alpha_{11}$ | $\alpha_{12}$ | $\alpha_{13}$ | $\alpha_{14}$ |
| | 0.3 | 0.92 | 2.05 | 1.7 | 0.87 | 1.5 | 1.7 | 2.2 | 2.8 | 4.7 | 0.94 | 1.03 | 1.085 | 2.35 | 2.65 |
| Day after beginning (day) | $\tau_0$ | $\tau_1$ | $\tau_2$ | $\tau_3$ | $\tau_4$ | $\tau_5$ | $\tau_6$ | $\tau_7$ | $\tau_8$ | $\tau_9$ | $\tau_{10}$ | $\tau_{11}$ | $\tau_{12}$ | $\tau_{13}$ | $\tau_{14}$ |
| | 0 | 6 | 31 | 55 | 63 | 77 | 95 | 121 | 126 | 138 | - | - | - | | |
| Day after beginning (Gd) | $T_1$ | $T_2$ | $T_3$ | $T_4$ | $T_5$ | $T_6$ | $T_7$ | $T_8$ | $T_9$ | | | | | | |

| (day) | 10 | 32 | 39 | 45 | 51 | 78 | 96 | 110 | 127 | | | | |
|---|---|---|---|---|---|---|---|---|---|---|---|---|---|
| Change of $S$ population | $\Delta \bar{S}_0^1$ | $\Delta \bar{S}_0^2$ | $\Delta \bar{S}_0^3$ | $\Delta \bar{S}_0^4$ | $\Delta \bar{S}_0^5$ | $\Delta \bar{S}_0^6$ | $\Delta \bar{S}_0^7$ | $\Delta \bar{S}_0^8$ | $\Delta \bar{S}_0^9$ | | | | |
| - | $-50\%$ | $-15\%$ | $-5\%$ | $-10\%$ | $-5\%$ | $+20\%$ | $-2\%$ | $-2\%$ | $-10\%$ | | | | |
| **Italy** | | | | | | | | | | | | | |
| Infection rate | $\alpha_0$ | $\alpha_1$ | $\alpha_2$ | $\alpha_3$ | $\alpha_4$ | $\alpha_5$ | $\alpha_6$ | $\alpha_7$ | $\alpha_8$ | $\alpha_9$ | $\alpha_{10}$ | $\alpha_{11}$ | $\alpha_{12}$ |
| $\times 10^{-8}$ $(day^{-1})$ | 3.75 | 8.75 | 6 | 4.75 | 5 | 5.6 | 6.52 | 7 | 8.3 | 7.3 | 8.5 | 10.5 | 10.6 |
| Day after beginning | $\tau_0$ | $\tau_1$ | $\tau_2$ | $\tau_3$ | $\tau_4$ | $\tau_5$ | $\tau_6$ | $\tau_7$ | $\tau_8$ | $\tau_9$ | $\tau_{10}$ | $\tau_{11}$ | $\tau_{12}$ |
| (day; date) | 0 | 10 | 20 | 35 | 45 | 55 | 60 | 65 | 70 | 80 | 90 | 125 | 133 |
| Day after beginning (Gd) | $T_1$ | $T_2$ | $T_3$ | $T_4$ | $T_5$ | | | | | | | | |
| (day; date) | 4 | 76 | 85 | 105 | 145 | | | | | | | | |
| Change of $S$ population | $\Delta \bar{S}_0^1$ | $\Delta \bar{S}_0^2$ | $\Delta \bar{S}_0^3$ | $\Delta \bar{S}_0^4$ | $\Delta \bar{S}_0^5$ | | | | | | | | |
| - | $-60\%$ | $+1\%$ | $+1\%$ | $+1\%$ | $+0.8\%$ | | | | | | | | |

### 4-3 Peaks and Repeating waves for CoViD-19

From the perspective of the control strategies and prevention measures, the peaks of the infectious population, and repetition of the epidemic waves during spreading, play a striking role [6,30,31]. The peaks of epidemics are related to the local and global extremums of the infected group, so the peaks occurred in time at which $dI(t)/dt = 0$ or equivalently for our model ($I(t_p) = \varepsilon E(t_p)/(\gamma + q + d_i)$). As can be seen, the characteristics of epidemic peaks (including amplitude and time gap and occurrence time) depend on the dynamical behaviors (in terms of modeling) as well as, the parameters of the model. In the past epidemics such as influenza-1918 multiple waves have been observed, for example, the United States experienced multiple waves in each three influenza pandemic that occurred in the 20[th] century (Spanish flu-1918, Asian flu-1957, and Hong-Kong flu-1968) [6,18,22,32]. There are several mechanisms that may lead to the repetition of waves of epidemics including the variations of transmission of the virus, virus mutation, heterogeneity (geography, demography and etc.) and loss of immunity [30,33]. In the present study, we propose a new mechanism that may lead to generating multiple waves, and it is the governmental large-scale measures in direction of the removing lockdown for quarantined people. Once again, consider the infected diagram of the active infected cases in Iran and Italy, suppose the government decides to reopen schools and universities, so a fraction of the population is transferred from the quarantine compartment to the direct subjection to the disease. Also, Figure (4) represents this effect on the progress of the epidemic in Iran and Italy, respectively. Note that, we have assumed that after reopening schools and universities (or in general any public places), the government will also apply mitigation measures such as social distancing, forcing people to wear a face mask etc. Therefore, the values of the transmission rates have been considered less than before the opening. Other tips that can be understood from Figure (4a) is that, although the mean coefficient of the transmission rate ($\alpha$) decreased from about late October (early November), reopening can lead to a great shock in the infectious individuals. Therefore, despite adopting mitigation measures of the transmission rate, quarantine removal

(even for a portion of the society) is dangerous, and should not be implemented, as much as possible. This phenomenon also verifies our proposed mechanism for producing multiple waves in the epidemic procedure. A notable tip about the presented results is that all of the conducted simulations have been utilized using assumed parameters, however, as far as possible, they have been considered logical, the main goal is only providing a vision of our proposed model for governmental decisions.

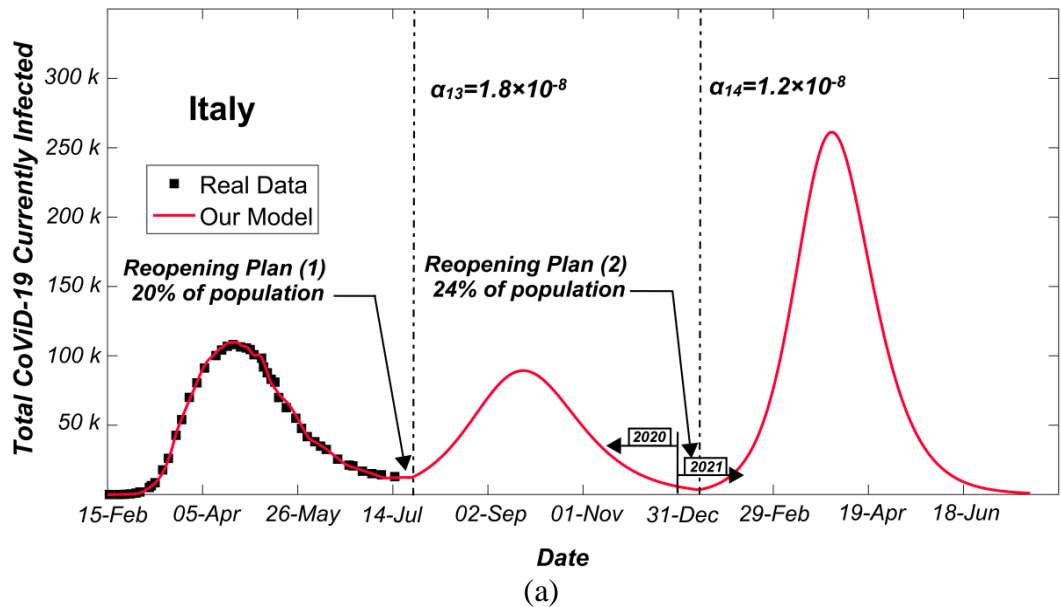

(a)

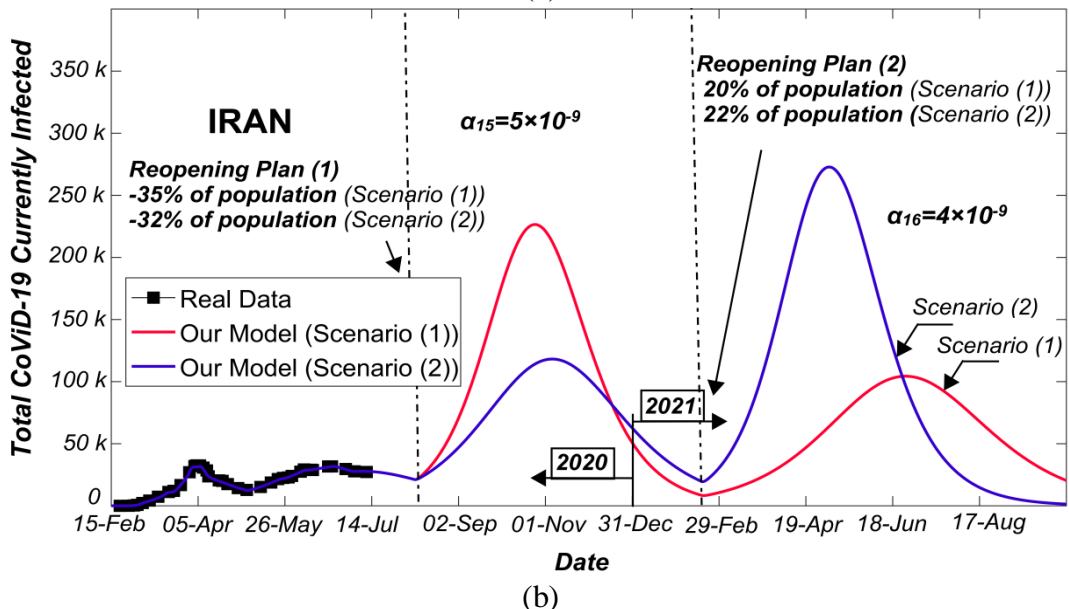

(b)

**Fig 4**: *The number of total active infected cases under the scenarios of the reopening for a) Italy and b) Iran. The prediction after 14$^{th}$ Jul has been accomplished using assumed coefficients and parameters ($\alpha_{15}$, $\alpha_{16}$ and etc.)* [20].

Figure (4b) shows the variations of the active infective cases for Iran under two different scenarios of the reopening and breaking of lockdown. As can be seen, the peak of the infected people increases when more individuals are free from quarantine. It should be mentioned that $2-3$ percent of the population (about $2\ million$ persons) has a notable efficacy on the amplitude of the peak. In addition, it can also be seen that the peak time occurs earlier, this result has an agreement with other conducted studies [34]. Another important finding associates with the response of society to the reopening plans at different decision times, Figure (5) illustrates this dependency. According to Figure (5) the time of the government decision (in the perspective of the reopening) effects on all peak properties including amplitude, time gap and occurrence time, in a way that, when the government is applied a reopening plan; whatever the number of the exposed and infectious individuals be lesser, the reopening plan leads to the smaller amplitude of the peak, lesser time gap as well as shift (forward) of the occurrence time. Therefore, our model does not recommend the breaking of the lockdown, when the number of the exposed people (consequently, infectious people) is still high.

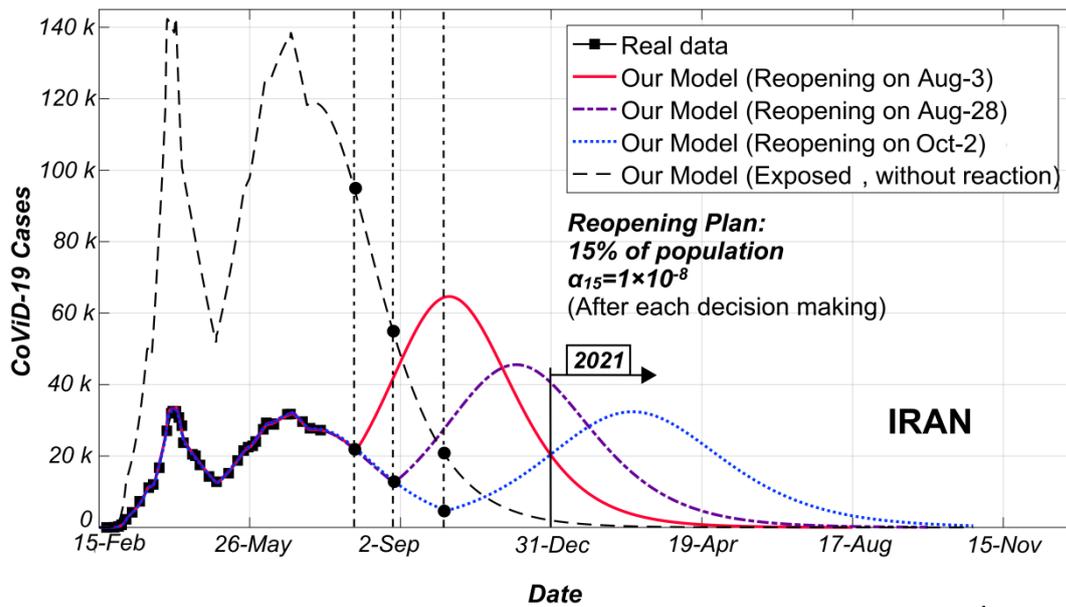

**Fig 5**: *The infectious cases for different reaction times The prediction after 14th Jul has been accomplished using assumed coefficients and parameters ($\alpha_{15}$ and etc.) [20].*

From the perspective of the healthcare capacity (e.g. Hospital beds and intensive care units (ICU)) we should note the peaks of the epidemic in a way that, in the peak times the system should be able to respond to all the patients [8]. The effects of the quarantine (isolation) rate of the infected cases ($q$) shown in figure (6). Figure (6) specified that the increasing temporary hospitals and CoViD-19 treatment (care) centers (increasing the value of the $q$) can reduce the number of infectious people especially in the peak time. Also, it can be found that, for the time-independent values of the isolation rate the time peaks do not change, this conclusion is logical due to the peak time governed equation ($I(t_p) = \varepsilon E(t_p)/(\gamma + q + d_i)$).

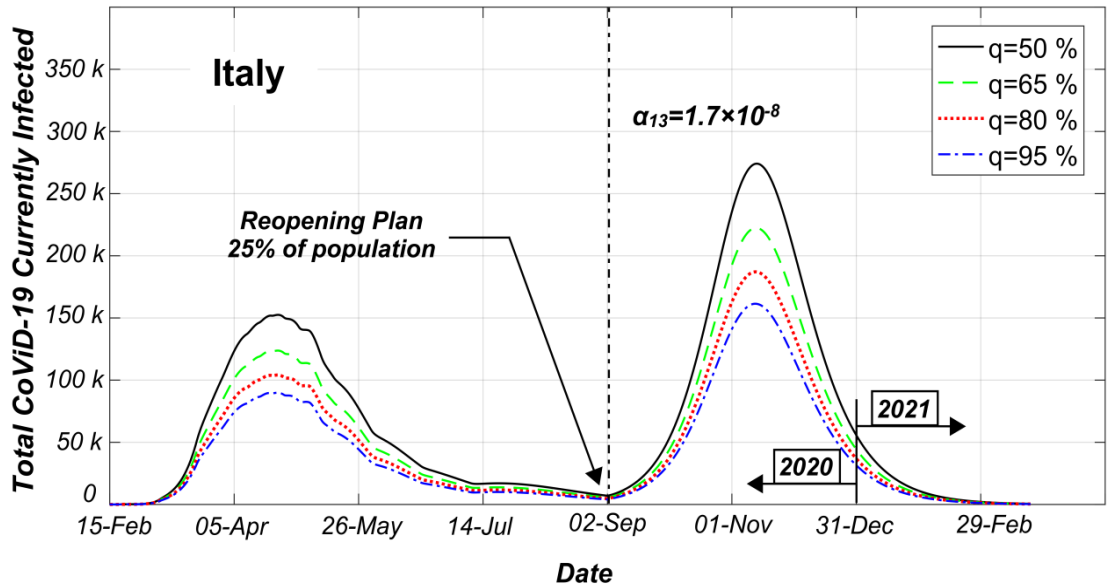

**Fig 6**: *The effects of the isolation rate.*

The environmental-related transmission of the disease is the next parameter that has an effect on the dynamical behavior of the spreading CoViD-19. Because of the novelty of our approach for environmental pollution-based transmission and lack of sufficient statistical data we use some desired pollution coefficient only in order to show its general effects (Figure (7)).

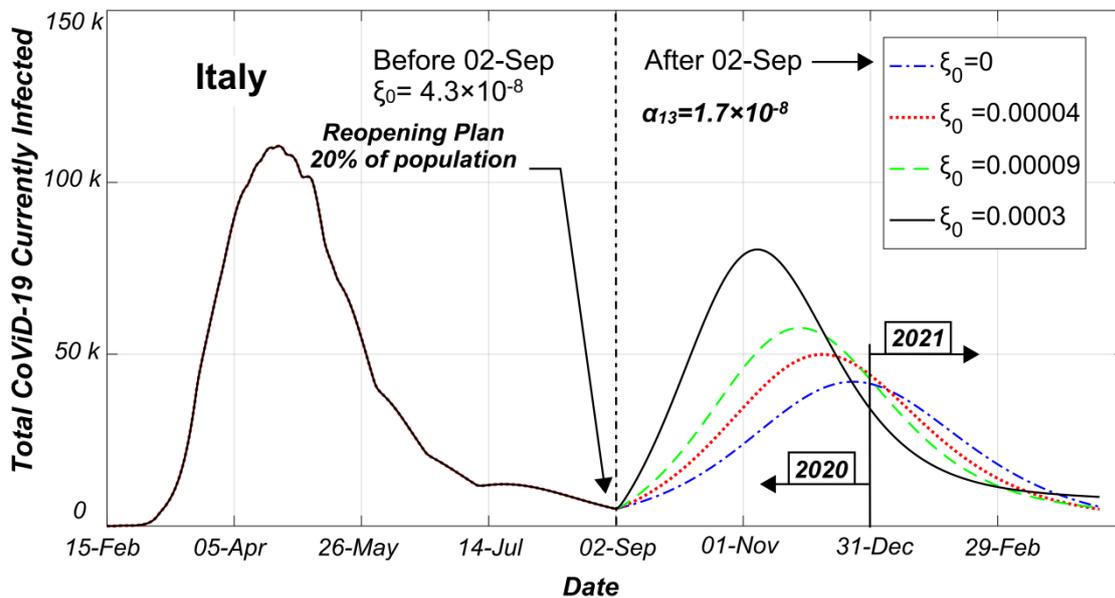

**Fig 7**: The variations of infectious people versus time for different pollution transmission coefficient (for after Sep 2).

Figure (7) displays that the number of infectious individuals increases when the environment is more polluted therefore, disinfection measures also to be considered as much as possible. Finally, the main question is "what can the government do after a reopening plan when the number of infectious cases is increased?" With regards to the proposed compartmental model, there are two basic answers, one reducing the possibility of the infection per contact by applying strict necessaries of personal healthcare and disinfection measures, and quarantining a fraction of the susceptible people once again [35]. These effects have been investigated in Figures (8a) and (8b), respectively.

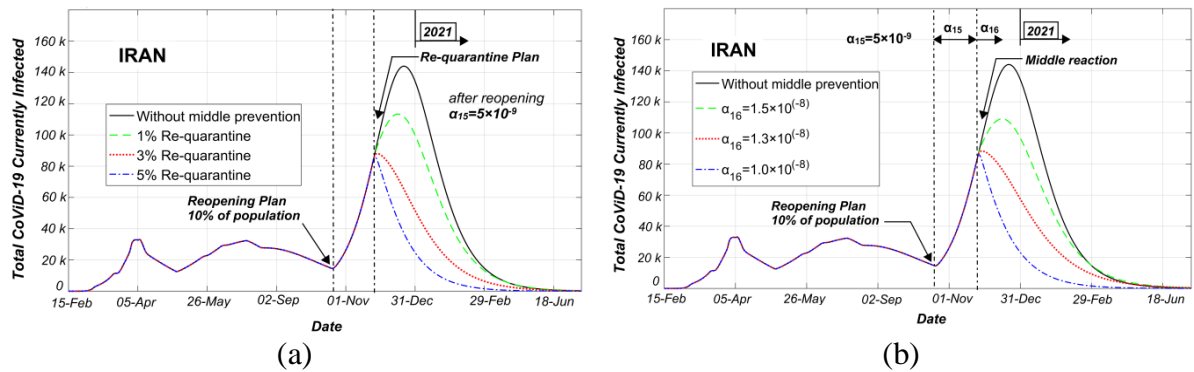

**Fig 8:** *The variations of infectious cases for two different mitigation measures a) re-quarantine plan and b) reducing transmission rate.*

Figure (8a) shows the effectiveness of the quarantine as a control measure. According to figure (8b), although reducing the transmission rate is very useful but may not be practically possible simply. It is worth mentioning that all of the parameters used in the simulation are assumed values but logical.

**4-4 Limitation and outlook**

Although, here the paper was focused on the CoViD-19 and the case studies were only for Iran and Italy, but, the proposed mathematical model can be utilized to forecasting each similar epidemic everywhere. Of course, extensive statistical data analysis should be used for accurate prediction and efficient providing optimum control strategies. The limitations of the proposed work can be classified as below:

(1) The model does not consider the male, female, children and other types of inhomogeneity dependent.

(2) The discussion about the basic reproductive number of our model in the presence of governmental actions should be improved in the future work.

Also as mentioned before, in order to the more efficient and realistic prediction of epidemic diseases such as CoViD-19 the model parameters and involved variables should be determined by statistical data.

**5 Conclusions**

In the present paper, a novel mathematical epidemic modeling was provided, which includes several basic elements such as governmental decisions (actions), environmental disease transmission, the behavioral response of the individuals, time-dependent model parameters etc. The governmental decision was applied in the model to consider the large-scale lockdown and reopening plan effects. In order to better describe the epidemic behavior and its progress pattern, some of the parameters were considered time-dependent. In addition, a new environmental-based transmission model was proposed to include the infection through the pollution places and surfaces. The equations governing the dynamical behavior of the system were solved using the 4$^{th}$-order Runge-Kutta method. Using the results from the solution of the model considering given coefficients and model parameters were fitted the spreading pattern of Iran and Italy. Also, the model can well predict the future of the epidemic, quantitatively, in the event that merged with accurate statistical data analysis. Under semi-logical assumptions of the simulation, the general conclusion and proposals based on the implemented analysis are summarized as the following.

- Taking into account the governmental decisions and environment-related pollution can better predict the pattern of the spread.
- The reopening plans may lead to multiple waves in the epidemics, even the transmission rate decreases.
- Even breaking the quarantine of 2-3 percent of the population of a common country can lead to highly shocks and waves in active infected cases.
- Reopening plans should be conducted with respect to the current situation of the epidemic since it would have existed more infected and exposed still should not break up lockdown even though the trend of the epidemic is decreasing.
- Sufficient re-quarantine and mitigation measures directly change the trend of the epidemics, therefore, in the re-ascent of the epidemic, they are recommended as much as possible.